\documentclass[12pt]{article}

\usepackage{epsfig}

\usepackage{amssymb}
\usepackage{amsmath}
\usepackage{amsfonts}

\usepackage{amsthm}

\theoremstyle{theorem}
\newtheorem{lem}{Lemma}[section]

\newtheorem{prop}[lem]{Proposition}

\theoremstyle{definition}

\newtheorem{pozn}[lem]{Remark}

\def\bp{\begin{proof}}
\def\ep{\end{proof}}

\def\be{\begin{equation}}
\def\ee{\end{equation}}
\def\ba{\begin{array}{c}}
\def\ea{\end{array}}

\def\ben{$$}
\def\een{$$}

\newcommand{\bea}{\begin{eqnarray}}
\newcommand{\eea}{\end{eqnarray}}

\begin{document}

\titlepage

\vspace{.35cm}

 \begin{center}{\Large \bf


Cryptohermitian Hamiltonians on graphs.

II. Hermitizations.

  }\end{center}

\vspace{10mm}

 \begin{center}

 {\bf Miloslav Znojil}

 \vspace{3mm}
Nuclear Physics Institute ASCR,

250 68 \v{R}e\v{z}, Czech Republic

{e-mail: znojil@ujf.cas.cz}

\vspace{3mm}

\end{center}

\vspace{5mm}

\section*{Abstract}

Non-hermitian quantum graphs possessing real (i.e., in principle,
observable) spectra are studied via their discretization. The
discretized Hamiltonians are assigned, constructively, an elementary
pseudometric and/or a more complicated metric. Both these
constructions make the Hamiltonian Hermitian, respectively, in an
auxiliary (Krein or Pontryagin) vector space or in a less friendly
(but more useful) Hilbert space of quantum mechanics.

\newpage

\section{Introduction}

In paper I \cite{I}, Schr\"{o}dinger equation $H \psi(x)=E \psi(x)$
with a non-Hermitian Hamiltonian $H=-\triangle + V \neq H^\dagger$
living on a non-tree toy-model graph $\mathbb{G}$ has been
considered.  We emphasized there that the unusual support
$\mathbb{G}$ of $H$ may find its multiple motivations in physics
using slightly smeared or non-local interactions.

In mathematical context we paid attention to the
meaning of the concept of the solvability of similar models. We
proposed that the latter concept finds its most natural
interpretation in the availability of the spectrum of energies in a
sufficiently transparent form. For this purpose we replaced the
``input" phenomenological Hamiltonian $H$ by an infinite family of
its discrete approximants $H^{(N)}$ and showed that and how this may
simplify the underlying secular equation via its factorization. In
this manner we were able to address the key problem emerging in
similar quantum graph models, namely, the problem of the
specification of the domain ${\cal D}$ of coupling constants for
which the whole spectrum remains real.

In our present continuation of paper I we intend to show that for
the couplings lying inside domain ${\cal D}$, all of the apparently
non-Hermitian Hamiltonians $H^{(N)}$ with $N \leq \infty$ may be
reinterpreted as Hermitian. In the first, preparatory step (cf.
section \ref{sec1}) we shall recall some basic references  and
summarize a few relevant details of quantum theory. We shall also
restrict our attention to the sufficiently elementary toy models
with a feasible specification of the domain ${\cal D}$ of the
reality (i.e., observability) of the energies.

In section \ref{sec2} we shall describe the construction of a
pseudo-metric ${\cal P}$ which obeys the relation
 \be
 H^\dagger\,{\cal P}-
 {\cal P}\,H=0\,
 \label{cryptogbe}
  \ee
and which will make our Hamiltonian pseudo-Hermitian, i.e., ${\cal
P}-$self-adjoint in a suitable {\em ad hoc} Krein or, more
precisely, Pontryagin space.

In section \ref{sec3} a Dyson's map $\Omega$ will be assumed to
exist inside ${\cal D}$, leading to an isospectral avatar
$\mathfrak{h} =\Omega\, H\,\Omega^{-1}$ of our Hamiltonian (cf.,
e.g., review \cite{SIGMA} for more details). By construction, the
latter operator proves self-adjoint in a certain ``paternal" Hilbert
space ${\cal H}^{(P)}$. In this context we shall remind the readers
that the latter representation space is, in practice, never used for
performing calculations. Its role is purely auxiliary. Its existence
just enables us to translate the Hermiticity of $\mathfrak{h}$ in
${\cal H}^{(P)}$ into the equivalent concept of the ``hidden
Hermiticity" of our original Hamiltonian in a unitarily equivalent
representation space ${\cal H}^{(S)}$ which is Hamiltonian-dependent
and which is constructed here {\em ad hoc}.

In the context of quantum physics the latter type of construction
has been first employed by Scholts, Geyer and Hahne \cite{Geyer}. It
has multiple merits. In Hilbert space ${\cal H}^{(S)}$, for example,
one can write relation $H = H^\ddagger$ (= hidden Hermiticity or
``cryptohermiticity" of $H$) where the new conjugate $H^\ddagger$ is
defined as an operator similar to $H^\dagger$. This similarity is
mediated by the metric operator defined as the product
$\Theta=\Omega^\dagger \Omega$ of Dyson's map with its conjugate. In
our final section \ref{sec4} we emphasize these connections and add
a few further relevant comments and commentaries.

\section{Discrete quantum graphs \label{sec1}}

In the abstract formalism of Quantum Mechanics the argument $x\in
{\cal Q}$ of wave function $\psi(x)$ may play the role of an
entirely formal variable. It need not necessarily be connected to a
point-particle position or momentum. Besides its more exotic but
still very traditional role of the time in quantum clocks \cite{Hoo}
(where the set ${\cal Q}$ still coincides with the real line) it may
even be chosen complex. For example, in the whole family of the so
called ${\cal PT}-$symmetric quantum models the most convenient set
${\cal Q}$ is being chosen in the form of a left-right symmetric
complex curve ${\cal C}(s)$ (cf. several recent reviews \cite{Carl}
of this innovative subject). In the so called quantum toboggans this
curve may even run over several Riemann sheets of the wave function
\cite{tobo}.

By its philosophy, paper I was closely related to the latter new
theoretical developments. It offered  a compact review of certain
potentially useful new family of quantum models where the set ${\cal
Q}$ is to be specified in the form of a suitable topologically
nontrivial (though still just real) graph $\mathbb{G}$. In the
present continuation of the short and sketchy text of paper I we are
going to complement the message. In particular, we intend to explain
how the exotic-looking quantum-graph models of paper I fit in the
standard textbook formalism of quantum mechanics.

In a compact summary of the contents of paper I we have to recall
that its mathematics was based on an $N-$point discretization ${\cal
Q}^{(N)}$ of the original graph-shaped ``kinematical" set ${\cal
Q}$. The promising physical implications of the less usual choice of
dynamics (i.e., of the interactions) has been motivated there by a
resulting short-ranged observational nonlocality of the models in
question. On constructive level the main attention of paper I (as
well as of our older paper \cite{web}) was devoted to the influence
of topological nontriviality of graphs ${\cal Q}^{(N)}$ (or,
ultimately \cite{Katka}, of their continuous-graph limits ${\cal
Q}^{(\infty)}$) upon the factorizations of the secular equations as
well as upon the reality and structure of the resulting spectra.

In the direction outlined in the conclusions of paper I we shall now
turn attention to the next task of the analysis. This task has two
aspects. On a purely formal level it lies just in a very
straightforward replacement of the ``false" Hilbert space ${\cal
H}^{(F)}$ by the ``standard" Hilbert space ${\cal H}^{(S)}$ (we use
the notation proposed in \cite{SIGMA}). This transition converts the
{\em manifestly non-Hermitian} Hamiltonian operator $H \neq
H^\dagger$ acting in ${\cal H}^{(F)}$ into its {\em manifestly
Hermitian} version acting in ${\cal H}^{(S)}$.

On a less formalistic level one has to emphasize that the initial
Hamiltonians $H$ are only considered at the so called physical
parameters, i.e., for the domains of couplings ${\cal D}^{(N)}$
which comply with the requirement that all the energies remain real
and also, for the sake of simplicity of the discussion,
non-degenerate. In this sense the main task of the users of
operators $H$ lies in the specification of the standard
Hilbert-space representation ${\cal H}^{(S)}$, i.e., in the {\em
explicit construction} of the above-mentioned metric operator
$\Theta=\Theta(H)$.

In the context of quantum theory on graphs, just the most elementary
samples of $H$ and $\Theta$ were shown obtainable in the tree-graph
elementary models of paper~\cite{fundgra}. In what follows we intend
to complement this construction by the less trivial samples of the
quantum graphs which combine the topological, mathematical
nontriviality of the corresponding supportive sets ${\cal Q}^{(N)}$
with the expected phenomenological nontriviality of the resulting
spectra of energies.

\subsection{Runge-Kutta-type discretizations }

In many papers dealing with concrete applications of Quantum Theory
the feasibility of practical model-building is based on a suitable
discretization of the spatial continuum (cf., e.g.,
Ref.~\cite{Katka} in this setting). In particular, in our papers
\cite{fund,disqw,disqwbe} we considered one-dimensional
Schr\"{o}dinger equations and replaced (i.e., approximated) the
underlying intervals of coordinates [say, $x \in (-\infty,\infty)$
or $x \in (-L,L)$] by suitable Runge-Kutta-type equidistant lattices
of points $x_k$ numbered by an integer subscript $k = \ldots, -1, 0,
1, \ldots$.

One of the most natural dynamical simulations of a {\em global},
long-ranged nonlocality is given by the replacement of the standard
straight-line {\em real} interval of $x \in (-L,L)$  in one
dimension (with either $L<\infty$ or $L=\infty$) by a tree-shaped
graph. The resulting planar, spatial or hyperspatial metric graph
can be assigned a toy-model Hamiltonian matrix.

A fairly large class of Hamiltonians admits not only a constructive
pseudo-Hermitization (i.e., a fully non-numerical reconstruction of
pseudometrics ${\cal P}={\cal P}(H)$) but also a constructive
pseudo-Hermitization (i.e., a fully non-numerical reconstruction of
positive definite metrics $\Theta=\Theta(H)$), in a large subdomain
of the domain of parameters where the spectrum is real. In the
absence of any interaction and for the finite number of the
Runge-Kutta lattice points such an idea leads merely to the discrete
version of the standard and solvable square-well model \cite{disqw}.
It may be perceived, say, as living on the linear discrete lattice
of $N=2K+1$ points,
 \be
 \ba
  \begin{array}{||c||}
 \hline
 \hline
 {\xi_{-K}}\\
 \hline
 \hline
 \ea \ {\xi_{-K+1}}\
 \ldots\   \xi_{-2}\  \xi_{-1}\
 \begin{array}{||c||}
 \hline
 \hline
 {\xi_0}\\
 \hline
 \hline
 \ea
 \  \xi_1\  \xi_2\  \ldots\ {\xi_{K-1}}\
   \begin{array}{||c||}
 \hline
 \hline
 {\xi_{K}}\\
 \hline
 \hline
 \ea  \\
  \ea\,.
 \label{VgstRKL}
 \ee
In terminology of Ref.~\cite{fundgra} the three marked points
$\xi_{-K}$, $\xi_{0}$ and $\xi_{K}$ may be reinterpreted as three
vertices of a discretized two-pointed-star graph $\mathbb{G}^{(2)}$.
In the same spirit one can reinterpret the corresponding quantum
square well as a discrete quantum graph \cite{exproc} in which the
dynamics of the system is controlled by $(2K+1)-$dimensional matrix
Schr\"{o}dinger equation
 \be
  \left[ \begin {array}{ccccccc}
   2&-1&0&\ldots&&\ldots&0
  \\
  -1
&2&-1&0&\ldots&\ldots&0
 \\0&-1&2&-1&\ddots&&\vdots
 \\
 \vdots&\ddots&\ddots&\ddots&\ddots&0&0
 \\
 {}&&
&-1&2&-1&0
 \\{}\vdots&&&\ddots&-1&2&-1
 \\{}0&\ldots&&\ldots&0&-1&2\\
 \end {array} \right]\,
 \left[ \begin {array}{c}
 \psi(\xi_{-K})\\
 \psi(\xi_{-K+1})\\
 \psi(\xi_{-K+2})\\
 \vdots\\
 \psi(\xi_{K-1})\\
 \psi(\xi_{K})\\
 \ea
 \right ]=E\,
 \left[ \begin {array}{c}
 \psi(\xi_{-K})\\
 \psi(\xi_{-K+1})\\
 \psi(\xi_{-K+2})\\
 \vdots\\
 \psi(\xi_{K-1})\\
 \psi(\xi_{K})\\
 \ea
 \right ]\,.
 \label{kinetie}
 \ee
%
\begin{figure}[h]                     
\begin{center}                         
\epsfig{file=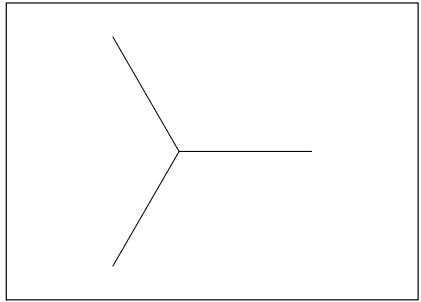,angle=270,width=0.4\textwidth}
\end{center}                         
\vspace{-2mm} \caption{Star-shaped graph $\mathbb{G}^{(q)}$ with
$q=3$.
 \label{figxxjed}}
\end{figure}
%
In a generalization of such a model one can replace the trivial
linear graph $\mathbb{G}^{(2)}$ by the three-pointed-star graph
$\mathbb{G}^{(3)}$ of Fig.~\ref{figxxjed} with three wedges and four
vertices. After a discretization such a graph coincides with the
T-shaped lattice
 \be
 \ba
  \begin{array}{||c||}
 \hline
 \hline
 {x_{N-2}}\\
 \hline
 \hline
 \ea\ {x_{N-5}}\
 \ldots\   x_5\  x_2\
 \begin{array}{||c||}
 \hline
 \hline
 {x_0}\\
 \hline
 \hline
 \ea
 \  x_3\  x_6\  \ldots\ {x_{N-4}}\  \begin{array}{||c|}
 \hline
 \hline
 {x_{N-1}}\\
 \hline
 \hline
 \ea  \\
 x_1\\
 x_4\\
 \vdots\\
  \begin{array}{||c||}
 \hline
 \hline
 {x_{N-3}}\\
 \hline
 \hline
 \ea
 \ea
 \label{YRKL}
 \ee
In Ref.~\cite{fundgra} we studied all the Schr\"{o}dinger equations
living on the $q-$pointed-star graphs at arbitrary $q$. In a move
beyond the traditional model-building framework we followed the
theory summarized in \cite{SIGMA} and introduced a new concept of a
non-Hermitian quantum graph. In our present paper we intend to move
to some topologically less trivial quantum graphs.

\subsection{Non-tree graphs}

%
\begin{figure}[h]                     
\begin{center}                         
\epsfig{file=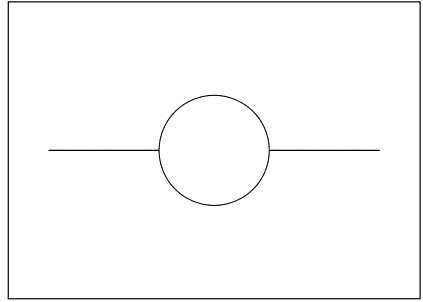,angle=270,width=0.4\textwidth}
\end{center}                         
\vspace{-2mm} \caption{One of the simplest non-tree graphs.
 \label{figujed}}
\end{figure}

 \noindent
In a purely numerical study~\cite{web} we replaced the topologically
trivial graph of Fig.~\ref{figxxjed} by its scale-dependent
loop-containing generalizations sampled by Fig.~\ref{figujed}. We
showed that similar quantum graphs still admit an efficient
application of the Runge-Kutta-type discretization techniques.

The simplest family of  discrete graphs contains the ``loop" part is
the form which is next to trivial,
 \be
 \ba
   \mbox{\ \ \ \ \ \ \ \
  }_\diagup\begin{array}{|c|}
 \hline
 {x_{0^+}}\\
 \hline
 \ea_\diagdown \\
  \ \begin{array}{||c||}
 \hline
 \hline
 {x_{-K}}\\
 \hline
 \hline
 \ea-\
 \ldots\   -\begin{array}{|c|}
 \hline
 {x_{-2}}\\
 \hline
 \ea-
 \begin{array}{||c||}
 \hline
 \hline
 \mbox{} x_{-1}\\
 \hline
 \hline
 \ea_\diagdown^\diagup
 \
 \begin{array}{c}
  \mbox{\ \ \
  \ \ \ \
 \ \ }\\
  \ea
 \ ^\diagdown_\diagup\begin{array}{||c||}
 \hline
 \hline
 \mbox{} x_1\\
 \hline
 \hline
 \ea - \begin{array}{|c|}
 \hline
 {x_{2}}\\
 \hline
 \ea -  \ldots -
   \begin{array}{||c||}
 \hline
 \hline
 {x_{K}}\\
 \hline
 \hline
 \ea  \\
   \mbox{\ \ \ \ \ \ \ \
 }^\diagdown\begin{array}{|c|}
 \hline
 {x_{0^-}}\\
 \hline
 \ea^\diagup  \\
  \ea
  \,.
 \label{VoooKL}
 \ee
Just the number $2K$ of points of the symmetric pair of the external
wedges can vary here. In the next step one can consider the lattices
 \be
 \ba
   \mbox{\ \ \ \ \ \
  }_\diagup\begin{array}{|c|}
 \hline
 {x_{U_1}}\\
 \hline
 \ea-
 \ldots -
 \begin{array}{|c|}
 \hline
 {x_{U_L}}\\
 \hline
 \ea_\diagdown \\
 \ \ \,
  \begin{array}{||c||}
 \hline
 \hline
 {x_{-K}}\\
 \hline
 \hline
 \ea   - \ldots -
 \begin{array}{||c||}
 \hline
 \hline
 \mbox{} x_{-1}\\
 \hline
 \hline
 \ea_\diagdown^\diagup
 \
 \begin{array}{c}
  \mbox{\ \ \
  \ \ \ \
  \ \
  \ \ \ \ \
  \ \ \ \ \ \ \
  \ \
  \ \ \ \
 \ \ }\\
  \ea
 \ ^\diagdown_\diagup\begin{array}{||c||}
 \hline
 \hline
 \mbox{} x_1\\
 \hline
 \hline
 \ea -  \ldots -
   \begin{array}{||c||}
 \hline
 \hline
 {x_{K}}\\
 \hline
 \hline
 \ea    \\
   \mbox{\ \ \ \ \ \
 }^\diagdown\begin{array}{|c|}
 \hline
 {x_{D_1}}\\
 \hline
 \ea-
 \ldots
 -\begin{array}{|c|}
 \hline
 {x_{D_L}}\\
 \hline
 \ea^\diagup  \\
  \ea
  \
 \label{V10KL3}
 \ee
containing the less trivial symmetric $2L-$point circular sublattice
which represents the loop and opens the possibility of mimicking the
the shape given by Fig.~\ref{figujed} in the limit of large $K,L \to
\infty$.

\subsection{Specific point-like interactions and their Hermiticity in disguise\label{ctvrta}}


In the spirit of Ref. \cite{SIGMA} the physical meaning and standard
probabilistic interpretation of {\em any} non-Hermitian Hamiltonian
$H\neq H^\dagger$ with real spectrum may be derived from its
Dyson-mapping-mediated Hermitian image
$\mathfrak{h}=\Omega\,H\,\Omega^{-1}$. As a rule, operator
$\mathfrak{h}$ is complicated by its form or counterintuitive by its
origin. Otherwise, there would be no reason for turning attention to
its isospectral-partner representation $H$. In this sense it is not
surprising that in the quantum-graph models of
Refs.~\cite{fundgra,fund} the Dyson's operators $\Omega$ appeared to
be fairly complicated.

Fortunately, the full knowledge of the latter operators is not too
often necessary in applications. The above-mentioned Hermiticity
condition
  \be
  \mathfrak{h}^{}=\Omega^{}\,H\
 \Omega^{-1}
  = \mathfrak{h}^\dagger\,.
  \label{edi}
  \ee
can be rewritten in the equivalent form of the Dieudonn\'{e}'s
\cite{Dieudonne} hidden-Hermiticity constraint
 \be
 H^\dagger
 =\Theta^{}\,H\,\Theta^{-1}\,,\ \ \ \ \ \ \ \
 \Theta=\Omega^\dagger\Omega\,.
 \label{cryptog}
  \ee
We quite often need not factorize $\Theta \to \Omega$. After all,
just the spectrum is usually sought and measured in experiments.

\section{The Hermitization of Hamiltonians in Krein space \label{sec2}}

\subsection{The choice of manifestly non-Hermitian interactions}

In a way complementing the recent illustrative
constructions~\cite{web,scatt} let us  turn attention to the
non-tree graph of Fig.~\ref{figujed} and to one of its most
elementary discrete versions or approximants
 \be
 \ba
   \mbox{\ \ \ \ \ \ \ \
  }_\diagup\begin{array}{|c|}
 \hline
 {x_{0^+}}\\
 \hline
 \ea_\diagdown \\
 \ \ \ \ \ \
  \begin{array}{||c||}
 \hline
 \hline
 {x_{-2}}\\
 \hline
 \hline
 \ea-
 \begin{array}{||c||}
 \hline
 \hline
 \mbox{} x_{-1}\\
 \hline
 \hline
 \ea_\diagdown^\diagup
 \
 \begin{array}{c}
  \mbox{\ \ \
  \ \ \ \
 \ \ }\\
  \ea
 \ ^\diagdown_\diagup\begin{array}{||c||}
 \hline
 \hline
 \mbox{} x_1\\
 \hline
 \hline
 \ea - \begin{array}{||c||}
 \hline
 \hline
 {x_{2}}\\
 \hline
 \hline
 \ea \,.\\
   \mbox{\ \ \ \ \ \ \ \
 }^\diagdown\begin{array}{|c|}
 \hline
 {x_{0^-}}\\
 \hline
 \ea^\diagup  \\
  \ea
  \ \
 \label{Voo6KL1}
 \ee
Once we endow its two central vertices $x_{-1}$ and $x_{1}$ with a
suitable non-Hermitian interaction we obtain the Hamiltonian
 \ben
 H(g,h;z)=\left[ \begin
 {array}{cccccc}
 2&-1-z&{}&{}&{}&{}\\
 -1+z& 3&-1-g&-1-h&{}&{}\\
 {}&-1+g&2&{}&-1+h&{} \\
 {}&-1+h&{}&2&-1+g&{}\\
 {}&{}&-1-h&-1-g &3&-1+z\\
 {}&{}&{}&{}&-1-z&2
 \end {array} \right]\,
 \een
in which the unperturbed free-motion matrix $H(0,0;0)$  is not too
different from its non-graph predecessor of Eq.~(\ref{kinetie}). It
is complemented by an elementary perturbation or interaction term
which manifestly violates the Hermiticity.

In the language of physics the assignment of the three-parametric
six-dimensional Hamiltonian $H(g,h;z)$ to the discrete graph
(\ref{Voo6KL1}) is directly inspired by Ref.~\cite{fund} where we
proposed that non-Hermitian interactions could simulate the presence
of an elementary length in the theory. We emphasized in~\cite{fund}
that the current trends \cite{Frikkie} of the introduction of the
fundamental length are different, relating this quantity directly to
certain hypothetical small anomalies in the geometry of the space or
space-time. The internal bubble in our graph (\ref{Voo6KL1}) can
very naturally be reinterpreted as one of such anomalies. This
returns us back to the mainstream literature where fundamental
length proved relevant, e.g. in field theory \cite{prva15}, in
string theory \cite{paty12}, in cosmology \cite{druha9} or in
astrophysics \cite{dvoj2}.

In our non-Hermitian models we can attribute the emergence of
non-localities not only to the small bubbles in the real-line graph
of coordinates but also to the smearing of space caused by the
manifestly non-Hermitain
interaction~\cite{fundgra,fund,scatt,Jones}. A very similar idea
appeared in preprint \cite{frej} where the concept of the
nonvanishing fundamental length found its theoretical origin in a
{\em combination} of the short-range spatial anomaly (viz.,
non-commutativity) {\em with} the smearing-effect of the
Dyson-mapping Hermitization $H \to \mathfrak{h}$. One can expect
model-building innovations of this type, say, in the context of the
solid state phenomenology (cf. a sample of activities in this
direction in Refs.~\cite{Jinbe}) or of the experimental optics
\cite{Muslimani} or, last but not least, of the recently revealed
possibility of the formation of microscopic spatial subdomains with
exotic properties in heavy-ion collisions \cite{Sumbera}.

\subsection{Pseudometric ${\cal P}$ in a toy model - a non-numerical construction
\label{sesta}}

One of the key sources of appeal of general non-Hermitian
Hamiltonians $H \neq H^\dagger$ possessing real spectra may be
traced back to letter \cite{BB} where Bender and Boettcher
demonstrated the mind-boggling reality of the spectra for the whole
one-parametric family of Hamiltonians $H=-\triangle +
V^{(\varepsilon)}(x) \neq H^\dagger$ containing complex potentials
$V^{(\varepsilon)}(x) = g^{(\varepsilon)}(x) x^2$ with
$g^{(\varepsilon)}(x)=({\rm i}x)^\varepsilon$. Tentatively they
assigned their observations to the so called ${\cal PT}-$symmetry of
their operators $H^{(\varepsilon)}$ and wave functions \cite{Carl}.

In the language of mathematics,  ${\cal P}$ and ${\cal T}$ need not
necessarily be just parity and time reversal operators as in
\cite{BB}. Moreover, the ${\cal PT}-$symmetry of $H$ should be
rephrased as the property written in the form of
Eq.~(\ref{cryptogbe}) which may be called ${\cal P}-$
pseudo-Hermiticity and which has already been studied, many years
ago, by mathematicians \cite{Dieudonne} as well as by physicists
\cite{BG}.

The most popular choice of ${\cal P}$ in the form of parity operator
enables one to treat Eq.~(\ref{cryptogbe}) as the  ${\cal
P}-$Hermiticity in a Krein space. We have shown in
Refs.~\cite{fundgra,fund} that for  graph-supported Hamiltonians the
search for a suitable operator ${\cal P}$ is much less trivial
(though still feasible) and that one only has to speak about the
${\cal P}-$Hermiticity of $H$ in a Pontryagin space. A {\em
complete} set of the linearly independent sparse-matrix operators
${\cal P}={\cal P}_n$ (compatible with Dieudonn\'{e}'s
Eq.~(\ref{cryptogbe})) has successfully been assigned to a selected
$H$ in \cite{fund}.


\begin{table}[h]
\caption{Dozen nontrivial matrix elements of Eq.~(\ref{cryptogbe})
after ansatz (\ref{anzad}).} \label{pexp4}
\begin{center}
\begin{tabular}{||c|c||c|c||}
\hline \hline
 {\rm position} &{\rm element to vanish}&{\rm position} &{\rm element to vanish}
 \\
 \hline
  \hline
                                  2&
                          -b + b z + a + a z&
                                  7&
                          -a - a z + b - b z\\
                                  9&
                   -{e} + {e} g - {u} + {u} h + b + b g&
                                  10&
                   -{u} + {u} g - {f} + {f} h + b + b h\\
                                  14&
                   -b - b g + {e} - {e} g + {u} - {u} h&
                                  17&
                   -c - c h + {e} - {e} h + {u} - {u} g\\
                                  20&
                   -b - b h + {u} - {u} g + {f} - {f} h&
                                  23&
                   -c - c g + {u} - {u} h + {f} - {f} g\\
                                  27&
                   -{e} + {e} h - {u} + {u} g + c + c h&
                                  28&
                   -{u} + {u} h - {f} + {f} g + c + c g\\
                                  30&
                          -d - d z + c - c z&
                                  35&
                          -c + c z + d + d z\\
  \hline
  \hline
\end{tabular}
 \label{tabone}
\end{center}
\end{table}

After we concentrate our attention to the discrete graph or lattice
(\ref{VoooKL}) with the growing number $K$ of grid points on each
external wedge, we have to imagine that besides the existing
explicit (and mostly numerical) analyses of the spectra (sampled,
say, in Refs.~\cite{I,web}), one can be interested in the
non-numerical aspects of these models. Thus, for our present
quantum-graph sample Hamiltonian $H(g,h;z)$ let us search for its
generalized parity via the most elementary nondiagonal ansatz
 \be
 {\cal P}=\left[ \begin {array}{cccccc} a&{}&{}&{}&{}&{}\\{}&b&{}&{}&{}
 &{}\\{}&{}&{\it {e}}&{\it {u}}&{}&{}\\{}&{} &{\it {u}}&{\it
 {f}}&{}&{}\\{}&{}&{}&{}&c&{}
 \\{}&{}&{}&{}&{}&d\end {array} \right]\,.
 \label{anzad}
 \ee
Its direct insertion in the Dieudonn\'{e}'s ``hidden-Hermiticity"
constraints (\ref{cryptogbe}) forms, in general, a set of 36 linear
equations which have to be satisfied by the 21 unknown (and, say,
real) matrix elements of the nondiagonal candidate ${\cal P}$ for
the pseudometric. In such a situation the 24 linear relations
(\ref{cryptogbe}) degenerate to identities and one is left just with
the twelve nontrivial right-hand matrix-element expressions
summarized in Table \ref{pexp4}.


\begin{table}[h]
\caption{Eight matrix elements of Eq.~(\ref{cryptogbe}) after the
elimination of $a$ and $d$.} \label{pexp6}
\begin{center}
\begin{tabular}{||c|c||c|c||}
\hline \hline
 {\rm position} &{\rm element to vanish}&{\rm position} &{\rm element to vanish}
 \\
 \hline
  \hline
                                 9&
                   -{e} + {e} g - {u} + {u} h + b + b g&
                                  10&
                   -{u} + {u} g - {f} + {f} h + b + b h \\
                                  14&
                   -b - b g + {e} - {e} g + {u} - {u} h&
                                  17&
                   -c - c h + {e} - {e} h + {u} - {u} g \\
                                  20&
                   -b - b h + {u} - {u} g + {f} - {f} h&
                                  23&
                   -c - c g + {u} - {u} h + {f} - {f} g \\
                                  27&
                   -{e} + {e} h - {u} + {u} g + c + c h&
                                  28&
                   -{u} + {u} h - {f} + {f} g + c + c g \\
  \hline
  \hline
\end{tabular}
 \label{tabtwo}
\end{center}
\end{table}

All of these expressions have to be made equal to zero by the
suitable choice of the unknown matrix elements in our ansatz for
${\cal P}$. As long as the 12 constraints of Table \ref{tabone} are
not all independent, we may employ the first one and the last one
and eliminate
 \ben
 a=\frac{1-z}{1+z}\,b\,,\ \ \ \ \ \ \ \ \ d=\frac{1-z}{1+z}\,c\,.
 \een
The resulting reduced system of eight equations is summarized in
Table \ref{pexp6}.

The final matrix solution ${\cal P}$ of Eq.~(\ref{cryptogbe}) can be
normalized, say, by the choice of
 \ben
 {u}=2\,.
 \een
This means that we arrive at the final four definitions of the
unknowns,
 \ben
 {e}=f=g+h\,,\ \ \ \ \ \ b=\frac{2+g-g^2-h-h\,g}{1+g}\,,
 \ \ \ \ \   c=\frac{2-g-h\,g+h-h^2}{1+h}\,.
 \een
One can immediately verify that  the resulting matrix ${\cal P}$ is
not positive definite. This means that it cannot be interpreted as a
metric in Hilbert space but merely as an indefinite metric in an
{\em ad hoc} specified and Hamiltonian-dependent Pontryagin space.


\subsection{The lattice with any size $N = 2K+2$\label{last}}

%

The symbolic-manipulation experience gained during the  construction
of ${\cal P}$ at $K=2$ can be extended to all the higher integers $K
\geq 3$ because we now know which matrix elements must be taken into
account in relations (\ref{cryptogbe}). This  facilitates the
determination of the matrix elements of ${\cal P}$ at any $K\geq 3$
via an amended ansatz which leads to the final formulae.

\begin{prop}
For our quantum-graph Hamiltonians $H(g,h;z)$ of  matrix dimension
$N=2K+2$, there exists a non-diagonal solution ${\cal P}$ of
Eq.~(\ref{cryptogbe}) with the following non-vanishing matrix
elements,
 \ben
 {\cal P}_{K+1,K+2}= {\cal P}_{K+2,K+1}=2\,,\ \ \ \ \
 {\cal P}_{K+1,K+1}= {\cal P}_{K+2,K+2}=g+h\,,
  \een
 \ben
 {\cal P}_{2,2}= {\cal P}_{3,3}=\ldots ={\cal P}_{K,K}=
 \frac{2+g-h-h\,g-g^2}{1+g}\,,
 \een
 \ben
  {\cal P}_{2K+1,2K+1}= {\cal P}_{2K,2K}=\ldots ={\cal
P}_{K+3,K+3}=
 \frac{2+h-g-g\,h-h^2}{1+h}\,
 \een
and
 \ben
  {\cal P}_{1,1}=\frac{1-z}{1+z}\,{\cal P}_{2,2}\,,
  \ \ \ \
  {\cal P}_{2K+2,2K+2}=\frac{1-z}{1+z}\,{\cal P}_{2K+1,2K+1}\,.
 \een
\end{prop}
 \bp
The proof by insertion is straightforward.
 \ep

The conclusions extracted form the formulae obtained at $K=2$ remain
unchanged.

\section{The Hermitization of Hamiltonians in Hilbert space \label{sec3}}

\subsection{Positive-definite metric $\Theta$  in a toy model\label{vuosma}}

The spectrum of our first nontrivial discrete quantum graph
(\ref{Voo6KL1}) with $K=2$ is easily evaluated. It proves composed
of the degenerate constant doublet $E^{(0)}_\pm=2$ complemented by
the quadruplet of certain coupling-dependent energies. It is worth
mentioning that once we reparametrize the couplings
$g=\gamma+\delta$ and $h=\gamma-\delta$, we obtain the two series of
energies
 \be
 E^{(2)}_\pm=E^{(2)}_\pm(\gamma,z)
 =\frac{5}{2}\pm \frac{1}{2}\,\sqrt{21-16\,{{\it \gamma}}^{2}-4\,z^2}\,,
 \label{fora}
 \ee
 \be
 E^{(1)}_\pm=E^{(1)}_\pm(\delta,z)
 =\frac{5}{2}\pm \frac{1}{2}\,\sqrt{5-16\,{{\it \delta}}^{2}-4\,z^2}\,
 \label{forbe}
 \ee
where one of the parameters is always absent. This property is also
exhibited by similar models at higher $K$s \cite{I}.

The domain ${\cal D}$ of the admissible couplings (i.e., of the
reality of the spectrum) will be rectangular at $z=0$. For example,
in the $K=2$ domain ${\cal D}$ we shall have $\gamma\in
(-\gamma_{(max)},\gamma_{(max)})$ and $\delta\in
(-\delta_{(max)},\delta_{(max)})$, with $\gamma_{(max)}=\pm
\sqrt{21/16}$ and $\delta_{(max)}=\pm \sqrt{5/16}$. Even when we
choose $z \neq 0$ we can still use formulae (\ref{fora}) and
(\ref{forbe}) for a close-form specification of the so called
``exceptional-point" boundary $\partial{\cal D}$ where the system
ceases to be Hermitizable.

The three-parametric nature of domain ${\cal D}$ makes its
description unnecessarily complicated, especially because the
parameters $g$ and $h$ both describe just a certain asymmetry of the
short-range interaction part of the Hamiltonian. For this reason let
us now simplify the system and contemplate solely its $\delta=0$
special cases with $H=H(g,g;z)$.

This restriction will certainly simplify the search for a metric
matrix $\Theta$ which must necessarily be real and positive. The
existence of such a matrix would immediately imply that some of the
eligible Dyson's matrices $\Omega$ may also very easily be defined
as the positive square roots of $\Theta$.

An important advantage of reduction $g=h$ is that it results in the
admissibility of $\Theta=\Theta^{(diagonal)}$. Even the naive
symbolic-manipulation software enables us to find
$\Theta^{(diagonal)}$ in the rather clumsy form
  \be
 \left[ \begin {array}{cccccc} {\frac { \left( 1+g- \left( 1+g
 \right) g \right)  \left( 1-z \right) }{ \left( 1+g \right)  \left(
 1 +z \right) }}&{}&{}&{}&{}&{}\\{}&{\frac {1+g- \left( 1+g
 \right) g}{1+g}}&{}&{}&{}&{}\\{}&{}&1+g&{}&{}&{}
 \\{}&{}&{}&1+g&{}&{}\\{}&{}&{}&{}&{\frac {1
 +g- \left( 1+g \right) g}{1+g}}&{}\\{}&{}&{}&{}&{}&{\frac { \left( 1+g-
 \left( 1+g \right) g \right)  \left( 1-z \right) }{
 \left( 1+g \right)  \left( 1+z \right) }}\end {array} \right]\,.
 \label{byhand}
  \ee
This result can and has to be simplified ``by hand".

Many non-diagonal matrices $\Theta=\Theta(H)$ may be constructed
using symbolic-manipulation methods. In order to illustrate this
possibility (which reflects just the well known ambiguity of the
metric \cite{Geyer}) we can even proceed non-numerically, by
combining our diagonal, parameter-free  metric $\Theta^{(diag.)}$ of
Eq.~(\ref{byhand}) with the pseudometric ${\cal P}$ of preceding
section \ref{sec2}. This yields the one-parametric family of the
nondiagonal candidates
 \be
 \Theta(\alpha)=\Theta^{(diag.)}+\alpha\,{\cal P}\,
 \label{alpha}
 \ee
for the metric. The necessary restriction
 \ben
 \frac{1+g}{1-g}\ >\ 2\,\alpha \ >\ -1
 \een
guarantees that the resulting matrix $\Theta(\alpha)$ is positive
definite and characterizes, therefore, the ultimate and sought
one=parametric family of the alternative Hilbert spaces ${\cal
H}^{(S)}(\alpha)$ of states with the non-equivalent but still
entirely standard probabilistic physical interpretation. Naturally,
these spaces differ by admitting {\em different} families of the
other, complementary observables \cite{Geyer}.

\subsection{Matrix Hamiltonians of any size $N = 2K+2$\label{ulast}}

%
%


Whenever one keeps all the free parameters inside domain ${\cal D}$
where, by definition, all the spectrum of energies is real, the
corresponding $N-$dimensional Hamiltonian matrix $H(g,h;z)$ may be
considered isospectral to its Hermitian partner
$\mathfrak{h}(g,h;z)$. According to the above-mentioned general
recipe, the correct probabilistic interpretation of the Hamiltonian
requires, therefore, that we find a solution $\Theta=\Theta(H)$ of
the Dieudonn\'{e}'s hidden-Hermiticity condition, i.e., of the
underdetermined linear set of equations (\ref{cryptog}).

In paragraph \ref{vuosma} we demonstrated that at $K=2$ one must be
a bit careful when using the computer-assisted symbolic
manipulations. For this reason we employed an amended code and
tested it on the $K=3$ problem with the simplified $g=h$ Hamiltonian
$H^{}(g,g;z)=$
 \ben
 \left[ \begin {array}{cccccccc} 2&-1-z&{}&{}&{}&{}&{}&{}
\\-1+z&2&-1&{}&{}&{}&{}&{}\\{}&-1&3&-1-g
&-1-g&{}&{}&{}\\{}&{}&-1+g&2&{}&-1+g&{}&{}
\\{}&{}&-1+g&{}&2&-1+g&{}&{}\\{}&{}&{}&-1-
g&-1-g&3&-1&{}\\{}&{}&{}&{}&{}&-1&2&-1+z
\\{}&{}&{}&{}&{}&{}&-1-z&2\end {array} \right]\,.
 \een
As the result we obtained the diagonal metric matrix with the
following non-vanishing matrix elements,
 \ben
 \Theta^{(diagonal)}_{1,1}=\Theta^{(diagonal)}_{8,8}=
 {\frac { \left( 1-z \right)  \left(
 1-g \right) }{1+z}}\,,
 \een
 \ben
 \ \ \ \
 \Theta^{(diagonal)}_{2,2}=\Theta^{(diagonal)}_{3,3}=
 \Theta^{(diagonal)}_{6,6}=\Theta^{(diagonal)}_{7,7}=
 1-g\,,
 \een
 \ben
 \ \ \ \ \
 \Theta^{(diagonal)}_{4,4}=\Theta^{(diagonal)}_{5,5}=
 1+g\,,
 \een
specified up to an arbitrary overall constant factor. The knowledge
of these formulae enables us to conclude that at least in the
square-shaped domain with $|z|<1$ and $|g|<1$ all of the energies of
the model remain real.

Although the similar recipe merely provides the sufficient condition
of the reality of the energies, we see that it may cover large
intervals of couplings. Similar quantum graphs with $g=h$ remain
also tractable non-numerically at the higher dimensions. Last but
not least, at the larger values of $K$ one  better appreciates the
difference between the dynamical roles of the  ``central" coupling
$g$ and the ``asymptotic" coupling~$z$~\cite{web}.  Even the ``first
nontrivial" $K=4$ Hamiltonian $H(g,g;z)$ represented by the
ten-dimensional matrix
 \ben
  \left[ \begin {array}{cccccccccc}
  2&-1-z&{}&{}&{}&{}&{}&{}&{}&{}
 \\-1+z&2&-1&{}&{}&{}&{}&{}&{}&{}\\
 {}&-1&2&-1&{}&{}&{}&{}&{}&{}\\
 {}&{}&-1&3&-1-g&-1-g&{}&{}&{}&{} \\
 {}&{}&{}&-1+g&2&{}&-1+g&{}&{}&{}\\
 {}&{}&{} &-1+g&{}&2&-1+g&{}&{}&{}\\
 {}&{}&{}&{}&-1-g&-1-g&3&-1&{}&{} \\
 {}&{}&{}&{}&{}&{}&-1&2&-1&{}\\
 {}&{}&{}&{}&{}&{}&{}&-1&2&-1+z\\
{}&{}&{}&{}&{}&{}&{}&{}&-1-z&2
\end {array}
 \right]
 \een
illustrates this comment and enables us to reveal the  general
pattern. This opens the way towards non-numerical characteristics of
Hamiltonians $H(g,g;z)$ at all the integers $K$ and dimensions
$N=2K+2$.

\begin{prop}
For the two-parametric subfamily of our quantum-graph Hamiltonians
$H(g,g;z)$ of matrix dimension $N=2K+2L$ with $L=1$  and with
parameters $g$ and $z$ in a domain ${\cal D}^{(K,L)}\subset {\cal
D}$, there exists a diagonal, positive solution $\Theta^{(K,L)}$ of
Eq.~(\ref{cryptog}) with the following non-vanishing matrix
elements,
 \ben
 \ \ \ \
 \Theta^{(K,L)}_{2,2}=\Theta^{(K,L)}_{3,3}=\ldots=\Theta^{(K,L)}_{K,K}=
  1-g\,,
 \een
 \ben
 \ \ \ \ \
 \Theta^{(K,L)}_{K+1,K+1}=\ldots=\Theta^{(K,L)}_{K+2L,K+2L}=
 1+g\,,
 \een
 \ben
 \ \ \ \
 \Theta^{(K,L)}_{K+2L+1,K+2L+1}=\ldots=\Theta^{(K,L)}_{2K+2L-1,2K+2L-1}=
  1-g\,
 \een
 and
 \ben
 \Theta^{(K,L)}_{1,1}=\Theta^{(K,L)}_{2K+2L,2K+2L}=
 {\frac { \left( 1-z \right)  \left(
 1-g \right) }{1+z}}\,.
 \een
 \label{onypr}
\end{prop}
 \bp
The proof by insertions is straightforward.
 \ep

 \begin{pozn}
We did not encounter any essential obstacles when we tried to
replace our Hamiltonians $H(g,g;z)$ living on the non-Hermitian
discrete quantum graph (\ref{VoooKL}) (where the inner loop contains
just four points, viz., $x_{\pm 1}$ and $x_{0^\pm }$) by their
generalizations living on similar discrete graphs containing $2L+2$
inner-loop points with $L=2, 3, \ldots$. In this sense also the
applicability and validity of Proposition \ref{onypr} may be
extended accordingly.
 \end{pozn}

 \begin{pozn}
Once we succeeded in finding {\em a} metric, our Hamiltonian
$H(g,g;z)$ may be declared Hermitian in the corresponding Hilbert
space ${\cal H}^{(S)}$. This means that its {\em spectrum} must {\em
necessarily be real} inside {\em all} the domain  ${\cal
D}^{(K)}\subset {\cal D}$ where the positive-definite metric exists.
Thus, the explicit construction of {\em some} $\Theta$ appears to be
a fairly efficient method of the rigorous proof of the reality of
the spectrum inside a subdomain of ${\cal D}$.
 \end{pozn}

\section{Discussion \label{sec4}}

We showed that the family of manifestly non-Hermitian discrete
quantum graphs admits not only its pseudo-Hermitization (i.e., the
construction of a certain generalized parity -- or pseudo-metric --
operator ${\cal P}$) but also its  Hermitization (i.e., an explicit
construction of metric $\Theta$ in some of the eligible ``physical"
or ``standard" Hilbert spaces ${\cal H}^{(S)}$).

Our present quantum-graph-building strategy can be perceived as an
ambitious realization of the innovative concept of spatial
nonlocality at short distances. In our present text a specific
non-tree graph realization of such a quantum structure has been
addressed as a certain first-step study aimed at a broader future
project. In our concrete, model-based analyses the structure of the
short-range anomalies has only been mimicked by a single small loop.
We believe that the transition to some more complicated graphs will
not lead to any technical complications in the future.

One of the traditional technical obstacles related to the study of
non-Hermitian Hamiltonians  can be seen in their rather difficult
perturbative (in)tractability \cite{Jones}. We circumvented this
obstacle by the discretization techniques. We have demonstrated that
this trick facilitated not only the analysis of spectra but also the
reconstruction of the metrics and pseudometrics.

Our text has been based on the presentation of Quantum Theory as
summarized in review~\cite{SIGMA}. In essence, every operator of an
observable quantity is assumed represented, {\em simultaneously}, in
several auxiliary Hilbert spaces ${\cal H}_j$. In contrast to the
current practice where only the unitarily equivalent Hilbert spaces
are considered (often, these Hilbert spaces are connected by
Fourier-type transformations), the innovated formalism requires that
the underlying one-to-one transformations $\Omega$  (often called
Dyson's mappings \cite{Geyer}) are not unitary so that, in general,
just one of the spaces (say, ${\cal H}_0={\cal H}^{(phys)}$) is
usually declared ``physical".

Historical origins of such an idea date back to pure mathematics
\cite{Dieudonne}. In physics, its repeated re-births were emerging
in perturbation theory \cite{BG,Caliceti},  in the theory of heavy
nuclei \cite{Geyer},  in field theory \cite{Kimball}, in quantum
cosmology \cite{alikg} etc. In all of these contexts one treats the
state-vector $\psi(x)$ and/or the generator of its time evolution
(i.e., Hamiltonian $H$) as quantities which only admit the correct
physical interpretation after a transition to the auxiliary space
${\cal H}^{(P)}$ {\em or rather} to its unitarily equivalent and,
generically, more friendly form ${\cal H}^{(S)}$ where the inner
product is defined via nontrivial metric $\Theta$.

The real popularity of such a formalism has been evoked by letters
\cite{BB,BBJ} where the mind-boggling ambiguity of the metric has
most efficiently been suppressed via an {\em ad hoc} requirement of
a charge times parity times time-reflection symmetry of the
Hamiltonian. This approach (carrying the historic \cite{BG} nickname
of ${\cal PT}-$symmetric quantum theory \cite{Carl}) should be
perceived as a {\em maximally nonlocal} formulation of the theory.
In Ref.~\cite{Jones} the identification of the ${\cal PT}-$symmetry
postulates with the extreme nonlocality of the operator of
coordinate is in fact based on reference to paper Ref.~\cite{cubic}
where such an ``infinite-range" characterization of the models in
question has been discovered in a slightly different context.

For this reason the ${\cal PT}-$symmetry-based recipe appeared
entirely unsuitable for extension from its current and natural
bound-state applications, say, to the sufficiently compact and
transparent description of the unitary scattering (cf. the more
detailed explanations in Ref.~\cite{Jonesdva}). In the literature
one finds two ways out of this difficulty. In one of them we simply
{\em turn attention to open systems}. The necessary (e.g.,
Feshbach's projection) techniques and their implementations have
been recently sampled in a long review paper \cite{Rotter} or in a
very short preprint on lattice models \cite{Jin}).

In our papers \cite{web,fund} the second possibility has been
investigated and endorsed. Now, we may formulate its extension ot
the problem of scattering on graphs as the open problem for imminent
analysis. The core of this new developments may be again expected to
lie in the replacement of the maximally nonlocal ${\cal
PT}-$symmetric recipe by {\em another guiding principle}. Requiring,
in essence, that the range of the smearing of coordinates caused by
the action of $\Theta$  (let us denote this fundamental length by
the usual symbol $\theta$) should be finite.

In the first preliminary application of this idea to bound states on
quantum graphs in Ref.~\cite{fundgra} the size of $\theta$ has been
left unspecified. In a proposal of continuation of these studies the
quantity $\theta$ should be understood as playing the role of a {\em
phenomenological} length which measures the range of the
not-too-nonlocal smearing of the measurements of coordinates
attributed to the non-Hermiticity of the tentative interaction
living on the graph.

\subsection*{Acknowledgements}

The support by the Slovak Research and Development Agency (contract
No. APVV-0071-06), by the Institutional Research Plan AV0Z10480505,
by the GA\v{C}R grant Nr. P203/11/1433 and by the M\v{S}MT ``Doppler
Institute" project LC06002  is acknowledged.


\end{document}